\documentclass[journal,10pt]{IEEEtran}

\usepackage{stfloats}
\usepackage[cmex10]{amsmath}
\usepackage{xmpaper,lzqsymbol}
\usepackage{amssymb,latexsym}
\usepackage{graphicx,subfigure}
\usepackage{subfigure}
\usepackage{color,cite,times}
\usepackage{epstopdf}
\usepackage{multirow}
\usepackage{algorithmic}
\usepackage[ruled,lined]{algorithm2e}
\usepackage{booktabs}
\usepackage{bbm} 
\usepackage{subfigure}
\usepackage{url}
\usepackage{fancyhdr}
\usepackage{cleveref}
\newcommand{\RNum}[1]{\uppercase\expandafter{\romannumeral #1\relax}}

\newcommand{\nc}{\newcommand}
\nc{\bsm}{\boldsymbol}
\nc{\mbs}{\mathbbmss}
\nc{\mbf}{\mathbf}

\begin{document}
\title{\fontsize{22pt}{28pt}\selectfont Hierarchical Passive Beamforming for Reconfigurable Intelligent Surface Aided Communications}
\author{Chang Cai, 
	Xiaojun Yuan, \IEEEmembership{Senior Member, IEEE,}
	Wenjing Yan, \IEEEmembership{Student Member, IEEE,}
	Zhouyang Huang,
	Ying-Chang Liang, \IEEEmembership{Fellow, IEEE,}
	and Wei Zhang, \IEEEmembership{Fellow, IEEE}
	\thanks{C. Cai, X. Yuan, W. Yan, Z. Huang, and Y.-C. Liang are with the Center for Intelligent Networking and Communications, 
	University of Electronic Science and Technology of China, Chengdu 611731, China (e-mail: caichang@std.uestc.edu.cn; xjyuan@uestc.edu.cn; wjyan@std.uestc.edu.cn; zhouyanghuangx@outlook.com; liangyc@ieee.org). 
	
	W. Zhang is with the School of Electrical Engineering and Telecommunications, University of New South Wales, Sydney NSW 2052, Australia (e-mail: wzhang@ee.unsw.edu.au). 
	}
}
\maketitle

\begin{abstract}
	In reconfigurable intelligent surfaces (RISs) aided communications,
	the existing passive beamforming (PB) design involves polynomial complexity in the number of reflecting elements, and thus is difficult to implement due to a massive number of reflecting elements.
	To overcome this difficulty, we propose a reflection-angle-based cascaded channel model by adopting the generalized Snell's law, in which the dimension of the variable space involved in optimization is significantly reduced, resulting in a simplified hierarchical passive beamforming (HPB) design.
	We develop an efficient two-stage HPB algorithm, which exploits the angular domain property of the channel, to maximize the achievable rate of the target user.
	Simulation results demonstrate the appealing performance and low complexity of the proposed HPB design.
\end{abstract}


\section{Introduction}
Reconfigurable intelligent surface (RIS) \cite{ChongwenHuang2019TWC}, a.k.a. intelligent reflecting surface (IRS) \cite{QingqingWu2019TWC} and large intelligent meta-surface (LIM) \cite{He2019Cascaded}, has been regarded as a promising new technology to revolutionize wireless networks.
Typically, RIS is a man-made planar array composed of a large number of sub-wavelength metallic or dielectric scattering elements, 
each of which is able to induce a certain amplitude and/or phase change to the electromagnetic waves \cite{ChongwenHuang2019TWC, QingqingWu2019TWC}.
By appropriately and dynamically adjusting the reflecting coefficient of each RIS element, impinging signals can be collaboratively combined and steered to desired directions to achieve a significant passive beamforming (PB) gain.

Recently, researches \cite{QingqingWu2019TWC, ChongwenHuang2019TWC, CunhuaPan2019TWC, HuayanGuo2020WSR, Ning2020SPGM} focusing on PB design are springing up.
The optimal/sub-optimal reflecting coefficients are derived based on various optimization objectives, e.g., energy efficiency maximization \cite{ChongwenHuang2019TWC}, transmit power minimization \cite{QingqingWu2019TWC}, and rate maximization \cite{CunhuaPan2019TWC,HuayanGuo2020WSR, Ning2020SPGM}.
However, due to the non-convexity of the PB design problems, the computational complexity rises polynomially as the number of RIS elements increases, causing unacceptable processing delay.
It becomes more severe in RIS-aided high-mobility communications since the tolerance of delay is much more stringent.
Moreover, the algorithms may fail to find a solution for dense deployment of large RISs, where hundreds of thousands reflecting elements are required to be properly adjusted.

Therefore, there is an urgent need to simplify the traditional PB design to meet the real-time signal processing requirement,
under the constraint that the performance loss is marginal.
In this letter, we tackle this challenge from the perspective of reducing the dimension of PB design variable space.
Inspired by the generalized Snell's law \cite{Nanfang2011Snell} stating that the reflection to any arbitrary directions can be achieved by setting the phases of each elements on a RIS to be arithmetic sequences,
we propose a reflection-angle-based cascaded channel model,
which dramatically reduces the number of optimization variables by adopting the structured phase shifts.
We then clarify that optimizing the phase difference between two adjacent elements (microscopic level design) and optimizing the reference phase of a RIS (macroscopic level design) correspond to adjusting the reflection angle and aligning the wavefront phases of the RIS-reflected beams, respectively,
which implies a hierarchical passive beamforming (HPB) design.
Aiming at maximizing the achievable user rate,
a low-complexity HPB algorithm is developed by dividing the RIS configuration design into a reflection angle adjustment stage and a wavefront phase alignment stage.
Numerical results further verify that the proposed HPB design reaps most of the performance gain of the traditional PB design and at the same time significantly reduces the computation time.

\section{Reflection-Angle-Based Cascaded Channel Model} \label{ChannelModeling}
\begin{figure}
	[t]
	\centering
	\includegraphics[width=.83\columnwidth]{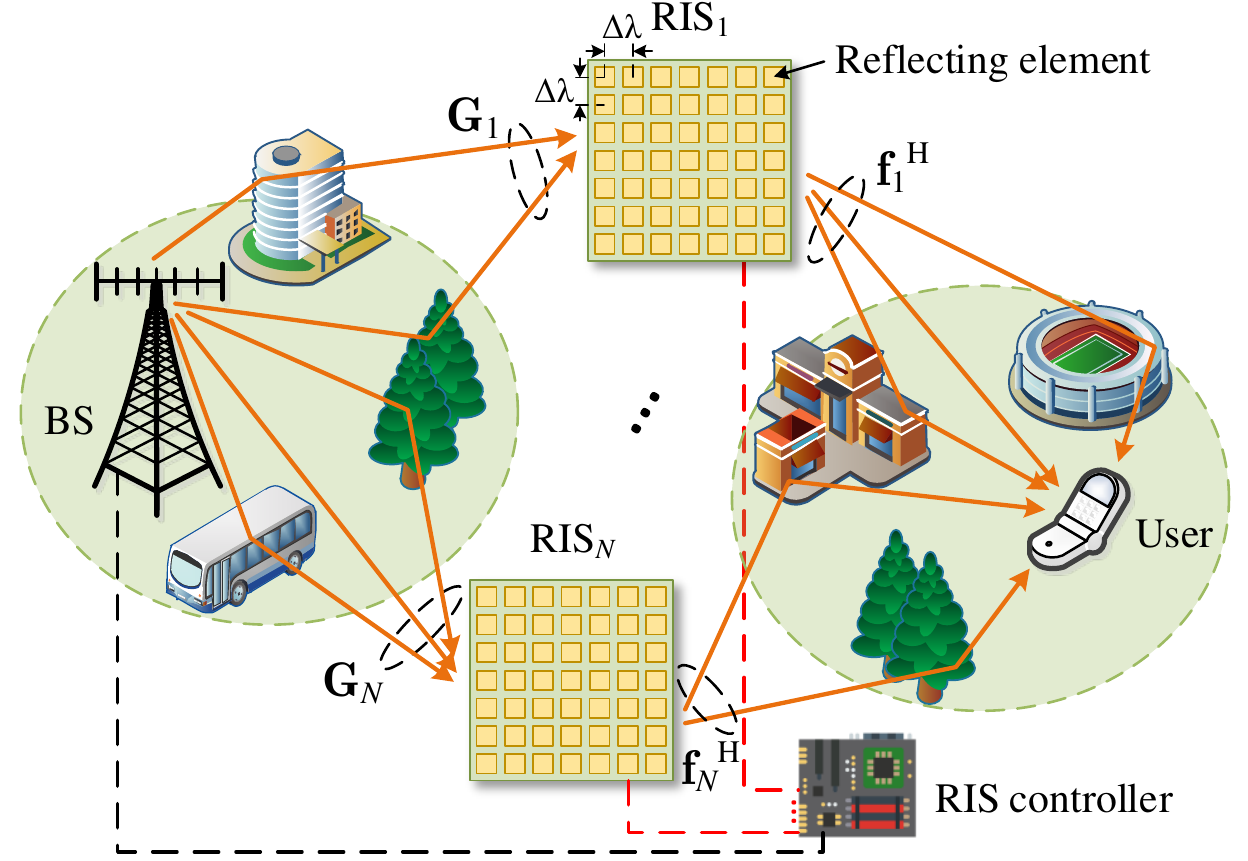}
	\vspace{-.6em}
	\caption{A RIS-aided cellular downlink communication system.}
	\label{HPB_MISO}
	\vspace{-1em}
\end{figure}
\begin{figure}
	[t]
	\centering
	\includegraphics[width=.7\columnwidth]{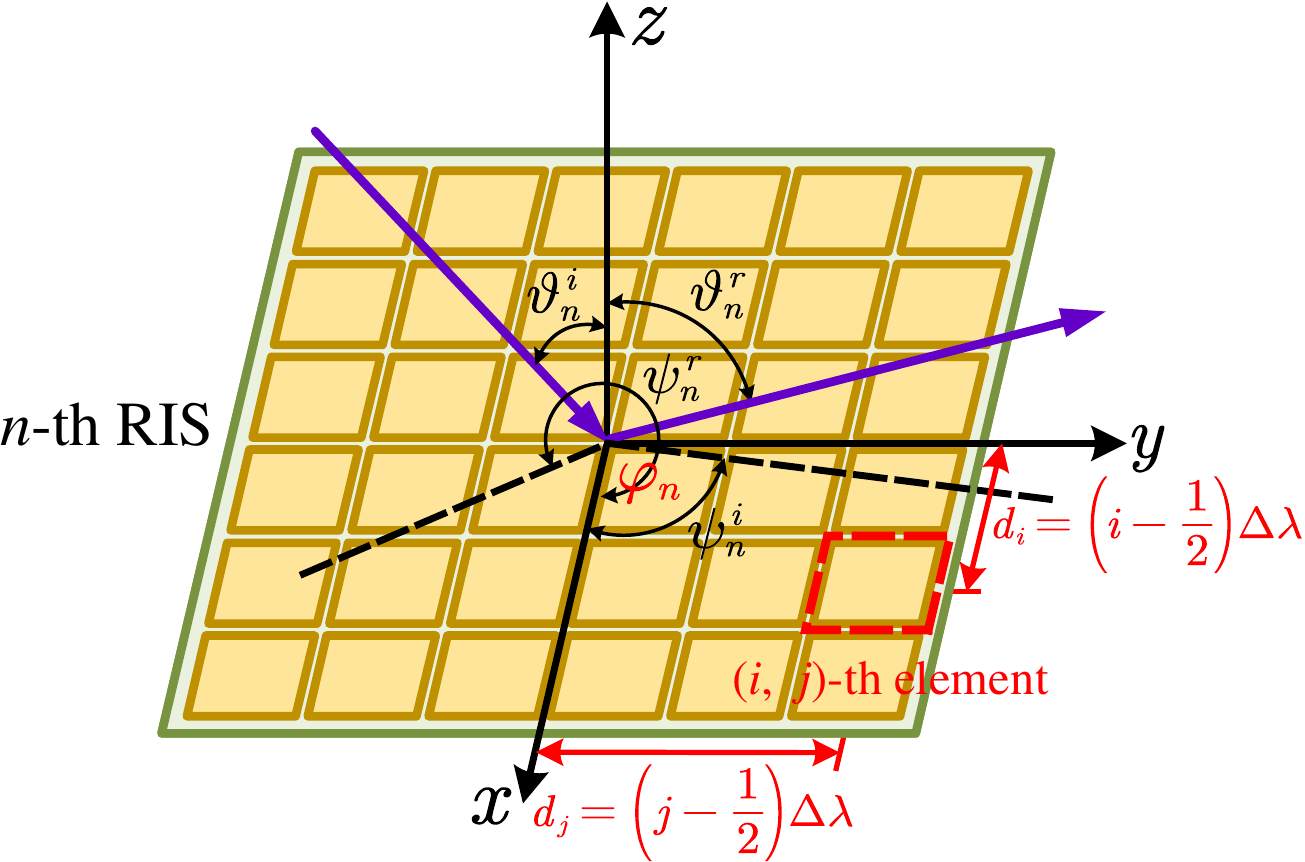}
	\vspace{-.6em}
	\caption{Illustration example of the generalized Snell's law.}
	\label{SystemModel}
	\vspace{-1em}
\end{figure}

As shown in Fig. \ref{HPB_MISO}, 
we consider a cellular downlink transmission
where one BS equipped with a half-wavelength uniform linear array (ULA) of $M$ isotropically radiating elements serves a single-antenna user.
The direct link between the BS and the user is blocked by obstacles.
In order to provide reliable communication services, $N$ co-located/distributed RISs are deployed to reflect the signals from the BS to the user.
We assume that each RIS is composed of $L^2$ reflecting elements with isotropic electromagnetic properties, arranged in an $L \times L$ uniform rectangular array (URA).
Taking the $n$-th RIS for illustration, as shown in Fig. \ref{SystemModel}, the RIS is placed in the $x$-$y$ plane of a Cartesian coordinate system, and the geometric center of the RIS is aligned with the origin of the coordinates.
The vertical and horizontal spacing between two adjacent elements is $\Delta \lambda$, where $\Delta$ is the scaling factor with respect to (w.r.t.) the signal carrier wavelength $\lambda$.
The center position of the $(i,j)$-th element is $\left( (i-\frac{1}{2})\Delta \lambda,(j-\frac{1}{2})\Delta \lambda,0 \right)$, where $i,j \in \left[ 1-\frac{L}{2},\frac{L}{2} \right]$, assuming that $L$ is an even number.
Before going through the details of the RIS-aided system, we first review the basic idea of the generalized Snell's law \cite{Nanfang2011Snell}, which will guide the RIS reflecting coefficients design later.

\subsection{Preliminary}
We consider an anomalous reflection scenario illustrated in Fig. \ref{SystemModel},
where $\vartheta_n^{i}, \psi_n^{i}, \vartheta_n^{r},$ and $\psi_n^{r}$ represent the elevation angle and the azimuth angle of the incident wave, the elevation angle and the azimuth angle of the reflected wave, respectively.
According to the generalized Snell's law \cite{Nanfang2011Snell}, to reflect an incident plane wave into another direction, breaking the specular reflection law ($\vartheta_n^{i} \neq \vartheta_n^{r}$ and/or $ \psi_n^{i} \neq  \psi_n^{r}+\pi $), the reflection phase of each element should depend linearly on the corresponding coordinate of both the $x$- and $y$-axes\footnote{
	It is known that a finite-size reflecting element may behave quite far from an ideal specular reflector.
	For example, it has been reported in \cite{Larsson2020Physics_WCL} that the beamwidth of the reflected wave from a $10\lambda \times 10\lambda$ RIS is about $10^{\circ}$.
	But even in this case, following the generalized Snell's law still guarantees that the main lobe of the beam is aligned with the desired direction.}.
Given the angle of incidence $(\vartheta_n^{i}, \psi_n^{i})$, the desired angle of reflection $(\vartheta_n^{r}, \psi_n^{r})$, and the reference phase of the origin of the coordinates $\varphi_n$,
the phase of the $(i,j)$-th element $\theta_{n,i,j}$ of the $n$-th RIS is obtained by \cite{Nanfang2011Snell}
\begin{align}
e^{\jmath \theta_{n,i,j}} = e^{\jmath \left( 2\pi \Delta \left( i-\frac{1}{2} \right) q_n^x 
+ 2\pi \Delta \left( j-\frac{1}{2} \right) q_n^y + \varphi_n \right)}, \label{SnellLaw}
\end{align}
where $q_n^x$ and $q_n^y$ denote respectively the phase gradient of the $x$- and $y$-axes of the $n$-th RIS, calculated as
\begin{align}
\mbf{q}_n \triangleq
\begin{bmatrix} q_n^x  \\  q_n^y \end{bmatrix}
=
\begin{bmatrix} \sin\vartheta_n^{r} \cos\psi_n^{r} + \sin\vartheta_n^{i} \cos\psi_n^{i} \\ \sin\vartheta_n^{r} \sin\psi_n^{r} + \sin\vartheta_n^{i} \sin\psi_n^{i} \end{bmatrix}. \label{q_n}
\end{align}
Eq. \eqref{q_n} reveals that the anomalous reflection from arbitrary $(\vartheta_n^{i}, \psi_n^{i})$ to arbitrary $(\vartheta_n^{r}, \psi_n^{r})$ can be achieved by setting $\mathbf{q}_n$ accordingly.
Besides, another degree of freedom (DoF) provided in \eqref{SnellLaw} is the reference phase $\varphi_n$ of the $n$-th RIS, which determines the wavefront phase of the reflected beam.
Thereby, instead of adjusting the phase of each reflecting element independently, the generalized Snell's law gives a paradigm to reduce the DoFs of PB design to three dimensions for each RIS, where $\mathbf{q}_n\in \mathbb{C}^{2\times 1}$ determines the direction of the reflected beam, and $\varphi_n\in \mathbb{C}^{1\times1}$ determines the wavefront phase.
Since the absolute values of both $\sin(\cdot)$ and $\cos(\cdot)$ are no larger than 1, $q_n^x, q_n^y \in [-2,2]$ always holds for any angles of incidence and reflection.
Moreover, due to the $2\pi$-periodicity of the phase shift, $q_n^x$ and $q_n^x + \frac{1}{\Delta}$ ($q_n^y$ and $q_n^y + \frac{1}{\Delta}$) yield the same RIS response function. 
Therefore, we have $q_n^x, q_n^y \in [-\bar{q},\bar{q}]$, where $\bar{q} = \min\left\{2, \frac{1}{2\Delta} \right\}$.



\subsection{Cascaded Channel Model}
The geometric channel model \cite{3GPP} is adopted for the individual BS-RIS and RIS-user links, which is based on the angle-of-departures (AoDs), angle-of-arrivals (AoAs), and the complex path gains of each path.
In this letter, we assume that the channel state information (CSI) is perfectly known at the BS.
Some compressive sensing techniques, such as the atomic norm minimization (ANM) method \cite{JiguangHe2020ANM_Hybrid} by exploiting the angular channel sparsity, can be applied to estimate the AoAs/AoDs and the corresponding path gains.
The multi-path channel $\mbf{G}_n \in \mathbb{C}^{L^2 \times M}$ from the BS to the $n$-th RIS can be expressed as
\begin{align}
\mbf{G}_n = \sum_{d=1}^{D_n} \alpha_{n,d}\mbf{a} (\vartheta_{n,d}^{\rm RIS},\psi_{n,d}^{\rm RIS} ) 
\mbf{b}^{\rm H} (\phi_{n,d}^{\rm BS}), \label{BS-RIS-Channel}
\end{align}
where $D_n$ denotes the number of resolvable paths;
$\alpha_{n,d}$ denotes the complex path gain of the $d$-th path;
$\vartheta_{n,d}^{\rm RIS}$ and $\psi_{n,d}^{\rm RIS}$ denote the corresponding elevation and azimuth AoAs, respectively;
and $\phi_{n,d}^{\rm BS}$ denotes the corresponding AoD.
The steering vector $\mbf{a} (\vartheta_{n,d}^{\rm RIS},\psi_{n,d}^{\rm RIS}) \in \mathbb{C}^{L^2 \times 1}$ is defined as
\begin{align}
\mbf{a} (\vartheta_{n,d}^{\rm RIS},\psi_{n,d}^{\rm RIS} )&=\mbf{a}^{x}(\vartheta_{n,d}^{\rm RIS},\psi_{n,d}^{\rm RIS} )\otimes \mbf{a}^{y}(\vartheta_{n,d}^{\rm RIS},\psi_{n,d}^{\rm RIS}), \label{URA-SteeringVector}
\end{align}
where $\otimes$ denotes the Kronecker product;
$\mbf{a}^{x}(\vartheta_{n,d}^{\rm RIS},\psi_{n,d}^{\rm RIS} ) \in \mathbb{C}^{L \times 1}$ and $\mbf{a}^{y}(\vartheta_{n,d}^{\rm RIS},\psi_{n,d}^{\rm RIS}) \in \mathbb{C}^{L \times 1}$ are given as follows: 
\begin{align}
\left[ \mbf{a}^{x}(\vartheta_{n,d}^{\rm RIS},\psi_{n,d}^{\rm RIS}) \right]_{\ell} &= \frac{1}{\sqrt{L}}e^{-\jmath 2 \pi \Delta (\ell-\frac{1}{2}) \sin\vartheta_{n,d}^{\rm RIS} \cos\psi_{n,d}^{\rm RIS}};\label{SteeringVector-ux}\\
\left[ \mbf{a}^{y}(\vartheta_{n,d}^{\rm RIS},\psi_{n,d}^{\rm RIS}) \right]_{\ell} &= \frac{1}{\sqrt{L}} e^{-\jmath 2\pi \Delta (\ell-\frac{1}{2})\sin\vartheta_{n,d}^{\rm RIS} \sin\psi_{n,d}^{\rm RIS}}, \label{SteeringVector-uy}
\end{align}
where $\jmath$ denotes the imaginary unit, and $\ell \in \left[1-\frac{L}{2},\frac{L}{2}\right]$.
The steering vector $\mbf{b} (\phi_{n,d}^{\rm BS}) \in \mathbb{C}^{M \times 1}$ is given by
\begin{align}
\left[\mbf{b} (\phi_{n,d}^{\rm BS})\right]_{m} = \frac{1}{\sqrt{M}}e^{-\jmath \pi (m-1) \sin\phi_{n,d}^{\rm BS}},
~~ m \in \left[1, M \right].
\end{align}

Similarly, the multi-path channel $\mbf{f}_n^{\rm H} \in \mathbb{C}^{1 \times L^2}$ from the $n$-th RIS to the user is modeled as
\begin{align}
\mbf{f}_{n}^{\rm H} = \sum_{k=1}^{K_n} \beta_{n,k}
\mbf{u}^{\rm H} (\theta_{n,k}^{\rm RIS},\phi_{n,k}^{\rm RIS}),
\end{align}
where $K_n$ denotes the number of resolvable paths;
$\beta_{n,k}$ denotes the complex path gain of the $k$-th path;
$\theta_{n,k}^{\rm RIS}$ and $\phi_{n,k}^{\rm RIS}$ denote the corresponding elevation and azimuth AoDs;
the steering vector $\mbf{u} (\theta_{n,k}^{\rm RIS},\phi_{n,k}^{\rm RIS}) \in \mathbb{C}^{L^2 \times 1}$ is defined in the same manner as $\mbf{a} (\vartheta_{n,d}^{\rm RIS},\psi_{n,d}^{\rm RIS} )$ in \eqref{URA-SteeringVector}.

Due to the severe product-distance path-loss and high attenuation, the signals reflected by the RISs more than once have negligible power and hence can be ignored\footnote{We note that there exists one line of research, e.g., \cite{You2021DoubleIRS}, that copes with cooperative double/multi-hop reflection problems, to further gain the benefits brought by deploying multiple RISs.}.
Moreover, by ignoring the direct link from the BS to the user, the received signal at the user can be expressed as
\begin{align}
y =  \sum_{n=1}^N \mbf{f}_{n}^{\rm H} \bsm{\Theta}_n \mbf{G}_n \mbf{w} s + n = \mbf{h}^{\rm H} \mbf{w} s + n, \label{Received-Signal}
\end{align}
where $\bsm{\Theta}_n = {\rm diag} \left\{ e^{\jmath \theta_{n,1-\frac{L}{2},1-\frac{L}{2}} }, \dots, e^{\jmath \theta_{n,\frac{L}{2},\frac{L}{2}}} \right\} \in \mathbb{C}^{L^2 \times L^2}$ denotes the diagonal phase shifts matrix for the $n$-th RIS, with reflecting amplitude being $1$;
$\mbf{w}$ denotes the transmit beamforming vector at the BS;
$s$ denotes the transmit data symbol;
$n$ denotes additive white Gaussian noise (AWGN) with zero mean and variance $\sigma^2$;
and $\mbf{h}^{\rm H} = \sum_{n=1}^N \mbf{f}_{n}^{\rm H} \bsm{\Theta}_n \mbf{G}_n$ denotes the cascaded BS-RIS-user channel.

\subsection{Reflection-Angle-Based Cascaded Channel Representation}
To reconfigure the wireless propagation environment, PB design problems have been widely studied in \cite{QingqingWu2019TWC, ChongwenHuang2019TWC, CunhuaPan2019TWC, HuayanGuo2020WSR, Ning2020SPGM} and the references therein. 
However, the aforementioned work considers that the phase shift of each RIS element is independently adjusted,
where a total of $NL^2$ variables should be jointly optimized. 
The complexity of solving such non-convex PB problems is sensitive to the size of the variable space and is in general very high.
For example, the involved complexity is $\mathcal{O} \left( (NL^2)^6 \right)$ via the semi-definite relaxation (SDR) method \cite{QingqingWu2019TWC}, which is prohibitively high especially for large surfaces consisting of thousands of reflecting elements. 
To meet the requirement of dense deployment of large RISs, we turn to explore the reflection property of each RIS, so as to significantly reduce the dimension of the PB variable space and thus simplify the optimization design.

The structured phase shifts defined by the generalized Snell's law in \eqref{SnellLaw} fully realize the reflection capability of a RIS.
By adopting the phase shift structure specified in \eqref{SnellLaw} for PB design, the variable space reduces from $NL^2$ to $3N$ for URA-shaped RISs, where only $q_n^x$, $q_n^y$ and $\varphi_n$ need to be customized for the $n$-th RIS.
Plugging \eqref{SnellLaw} into the cascaded channel $\mbf{h}^{\rm H}$, we have\footnote{The steering vectors $\mbf{u} (\theta_{n,k}^{\rm RIS},\phi_{n,k}^{\rm RIS})$, $\mbf{a} (\vartheta_{n,d}^{\rm RIS},\psi_{n,d}^{\rm RIS} )$, and $\mbf{b} (\phi_{n,d}^{\rm BS})$ are abbreviated to $\mbf{u}_{n,k}$, $\mbf{a}_{n,d}$, and $\mbf{b}_{n,d}$ in the sequel of this paper.}
\begin{align}
\mbf{h}^{\rm H} &= \sum_{n=1}^N \left( \sum_{k=1}^{K_n} \beta_{n,k} \mbf{u}^{\rm H}_{n,k} \right)
\bsm{\Theta}_n
\left( \sum_{d=1}^{D_n} \alpha_{n,d}\mbf{a}_{n,d} \mbf{b}^{\rm H}_{n,d} \right)
 \nonumber \\
&= \sum_{n=1}^N \sum_{k=1}^{K_n}\sum_{d=1}^{D_n} \alpha_{n,d} \beta_{n,k} p_{n,k,d} \mbf{b}_{n,d}^{\rm H},\label{CascadedChannel-Sinc-Sinc}
\end{align}
where
\begin{align}
p_{n,k,d} &= \left(\mbf{u}_{n,k}^x \otimes \mbf{u}_{n,k}^y\right)^{\rm H}
\bsm{\Theta}_n
\left(\mbf{a}_{n,d}^x \otimes \mbf{a}_{n,d}^y\right)
\nonumber \\
&= \sum_{i=1-\frac{L}{2}}^{\frac{L}{2}}\sum_{j=1-\frac{L}{2}}^{\frac{L}{2}}
\big[\mbf{u}_{n,k}^x\big]_{i}^{*} \big[\mbf{u}_{n,k}^y\big]_{j}^{*} 
e^{\jmath \theta_{n,i,j}}
\big[\mbf{a}_{n,d}^x\big]_{i} \big[\mbf{a}_{n,d}^y\big]_{j}  \nonumber \\
&= \frac{e^{\jmath \varphi_n}}{L^2}  \sum_{i=1-\frac{L}{2}}^{\frac{L}{2}} e^{-\jmath 2\pi \Delta (i-\frac{1}{2})s_{n,k,d}^{x}}
\sum_{j=1-\frac{L}{2}}^{\frac{L}{2}} e^{-\jmath 2\pi \Delta (j-\frac{1}{2})s_{n,k,d}^{y}}  \nonumber \\
&\overset{(a)}{=} e^{\jmath \varphi_n} \frac{{\rm sinc}\left( \Delta L s_{n,k,d}^x\right)}{{\rm sinc}\left(\Delta s_{n,k,d}^x\right)}
\frac{{\rm sinc}\left(\Delta L s_{n,k,d}^y\right)}{{\rm sinc}\left(\Delta s_{n,k,d}^y\right)},\label{direction}
\end{align}
with
\begin{align}
s_{n,k,d}^{x} &  =  \sin\theta_{n,k}^{\rm RIS} \cos\phi_{n,k}^{\rm RIS} + \sin\vartheta_{n,d}^{\rm RIS} \cos\psi_{n,d}^{\rm RIS} - q_{n}^{x}; \label{Sinc-Variable-x}\\
s_{n,k,d}^{y} & = \sin\theta_{n,k}^{\rm RIS} \sin\phi_{n,k}^{\rm RIS} + \sin\vartheta_{n,d}^{\rm RIS} \sin\psi_{n,d}^{\rm RIS} - q_{n}^{y} .\label{Sinc-Variable-y}
\end{align}
In \eqref{direction}, step $(a)$ can be derived by applying the sum of the geometric progression.
As expected, \eqref{direction} shows the beam pattern of a finite-size RIS reflector.
We rewrite \eqref{CascadedChannel-Sinc-Sinc} in a more compact form as
\begin{align}
\mbf{h}^{\rm H} =  \mbf{v}^{\rm H} \mbf{H}, \label{Cascaded-Channel}
\end{align}
where
$\mbf{v} = \big[ e^{\jmath \varphi_1} , \dots, e^{\jmath \varphi_N}\big]^{\rm H}$,
$\mbf{H} = \big[\mbf{r}^{\rm H}_1 \mbf{B}_{1}, \dots, \mbf{r}^{\rm H}_N \mbf{B}_{N} \big]$,
$\mbf{B}_{n} = \big[ \mbf{b}_{n,1}, \dots, \mbf{b}_{n,D_n} \big]^{\rm H}$,
$\mbf{r}_n = \big[r_{n,1}, \dots, r_{n,D_n}\big]^{\rm H}$,
and $r_{n,d} = \alpha_{n,d} \sum_{k=1}^{K_n} \beta_{n,k} \left| p_{n,k,d} \right|$.

It is observed that the phase shift component in \eqref{SnellLaw} is divided into two separate parts in \eqref{Cascaded-Channel}, 
i.e., $\mbf{Q} = \left[\mbf{q}_1, \dots, \mbf{q}_N\right]$ inside $\mbf{H}$
and $\mbf{v}$ outside $\mbf{H}$.
The wireless channel thus can be reconfigured from two aspects.
Firstly, from the element level, the reflection angle of each RIS can be controlled by adjusting the phase difference, i.e., $\mbf{Q}$, between two adjacent elements, which can be regarded as a microscopic PB design.
Secondly, from the RIS level, the reference phase of each RIS, i.e., $\varphi_n$, is adjusted to align the phase of the reflected beam arrived at the user, which can be regarded as a macroscopic PB design.
This leads to a reflection-angle-based HPB problem, as formally stated in the following section.

\section{Problem Statement}\label{ProblemStatement}
We aim to maximize the achievable rate of the user subject to the maximum transmit power constraint at the BS.
In the single-user setup in this letter, the objective reduces to the maximization of the received signal power.
Thus, the problem can be formulated as
\begin{subequations}
	\begin{align} \label{P1}
	\max_{\mbf{w}, \mbf{Q}, \mbf{v}} \quad &  \left| \mbf{v}^{\rm H} \mbf{H} \mbf{w} \right|^2 \\ 
	\operatorname{ s.t. } \quad
	& \left\| \mathbf{w} \right\|^{2} \leq p,\\
	& \left| v_n \right| = 1, ~~\forall n, \label{P1-vn}\\
	& q_n^x, q_n^y \in [-\bar{q},\bar{q}], ~~\forall n,\label{P1-qn}
	\end{align}
\end{subequations}
where $\mbf{Q}$ and $\mbf{v}$ are the HPB variables. 
Maximum-ratio transmission (MRT) is applied to obtain the optimal transmit beamforming solution, i.e., $\mbf{w}^{\star} = \sqrt{p} \frac{\mbf{H}^{\rm H} \mbf{v}}{\left\| \mbf{v}^{\rm H} \mbf{H} \right\| }$.
By substituting $\mbf{w}^{\star}$ into the above problem, it can be simplified to the following equivalent problem:
\begin{align}
	\max_{ \mbf{Q}, \mbf{v} } \quad f\left( \mbf{Q}, \mbf{v} \right) \triangleq \left\| \mbf{v}^{\rm H} \mbf{H} \right\|^2 \quad\quad
	\operatorname{ s.t. } \quad
	\eqref{P1-vn},\eqref{P1-qn}. \label{Problem_vQ}
\end{align}

\section{Hierarchical Passive Beamforming Design} \label{Algorithm}
In general, alternating optimization (AO) based algorithm is needed to recursively optimize $\mbf{Q}$ and $\mbf{v}$ until convergence.
The iteration involved may be quite time-consuming.
Moreover, it can be observed from \eqref{CascadedChannel-Sinc-Sinc} to \eqref{Cascaded-Channel}
that $f\left( \mbf{Q}, \mbf{v} \right)$ is highly non-convex w.r.t. $\mbf{Q}$ due to the involvement of sinc functions.
Thus, the gradient descent algorithms for optimizing $\mbf{Q}$ are likely to be stuck in local optimum.
As such, we propose a low-complexity two-stage algorithm, termed HPB-Strongest-Path-Pairing (HPB-SPP), to solve problem (16) sub-optimally by exploiting the angular domain property.
The details of the algorithm are specified as follows.

\begin{algorithm} [t]
	\caption{HPB-SPP Algorithm}
	\label{alg:SCA}
	\KwIn{Maximum number of iteration $I_{\rm sca}$; convergence criterion $\epsilon_{\rm sca}$.}
	\textbf{Initialize:}{ $\mbf{v}^0$; $t \leftarrow 1$.} \\
	\underline{\textbf{Stage \RNum{1}:}}\\
	\begin{itemize}
		\item Obtain the phase gradient $\mathbf{Q}$ according to \eqref{q_design}.
	\end{itemize}
	\underline{\textbf{Stage \RNum{2}:}}\\
	\begin{itemize}
		\item Update $\mbf{v}$ according to \eqref{v_update}.
		\item Let $t \leftarrow t+1$ and repeat \textbf{Stage \RNum{2}} until the relative decrease is less than $\epsilon_{\rm sca}$ or $I_{\rm sca}$ is attained.
	\end{itemize}
	\KwOut{$\mbf{Q}, \mbf{v}$.}
\end{algorithm}

\subsubsection{Stage \RNum{1}}
In typical wireless environment,
the number of distinguishable paths in the angular domain is much smaller than that of the reflecting elements.
Besides, a small portion of distinguishable paths occupy the majority of the channel power gain \cite{3GPP}.
Motivated by this, in the first stage, we propose to design $\mbf{q}_n (\forall n)$ such that the signals coming from the strongest path of the BS-RIS$_n$ channel are steered right towards the target user through the strongest path of the RIS$_n$-user channel.
Let $d_n^{\star}$ and $k_n^{\star}$ denote respectively the strongest path of the individual BS-RIS$_n$ and RIS$_n$-user links,
mathematically given by
\begin{equation}
	d_n^{\star} = \mathrm{arg} \max_{d=1,\cdots,D_n} \left| \alpha_{n,d} \right|;~~
	k_n^{\star} = \mathrm{arg} \max_{k=1,\cdots,K_n} \left| \beta_{n,k} \right|.
\end{equation}
Then, the gradient $\mbf{q}_n$ is set by applying the generalized Snell's law:
\begin{align}
\mbf{q}_n =
\begin{bmatrix} \sin\theta_{n,k_n^{\star}}^{\rm RIS} \cos\phi_{n,k_n^{\star}}^{\rm RIS} + \sin\vartheta_{n,d_n^{\star}}^{\rm RIS} \cos\psi_{n,d_n^{\star}}^{\rm RIS} \\ \sin\theta_{n,k_n^{\star}}^{\rm RIS} \sin\phi_{n,k_n^{\star}}^{\rm RIS} + \sin\vartheta_{n,d_n^{\star}}^{\rm RIS} \sin\psi_{n,d_n^{\star}}^{\rm RIS} \end{bmatrix}, ~\forall n. \label{q_design}
\end{align}

\subsubsection{Stage \RNum{2}}
In the second stage, the design of $\mbf{v}$ is considered to align the wavefront phases of the beams that reflected by each RIS, to maximize the received signal power.
For well designed $\mbf{Q}$, the problem in \eqref{Problem_vQ} reduces to
\begin{align}
\max_{\mbf{v}} \quad  f\left( \mbf{v} \right) \quad\quad\quad
\operatorname{ s.t. } \quad \eqref{P1-vn} \label{problem_v}.
\end{align}
This is a non-convex quadratically constrained quadratic program (QCQP) problem and can be iteratively solved by the successive convex approximation (SCA) algorithm.
For given $\mbf{v}^{t-1}$ at the $t$-th iteration, we obtain from convexity that
\begin{align}
f(\mbf{v}) \geq 2 \mathrm{Re} \left\{ (\mbf{v}^{t-1})^{\rm H} \mbf{H} \mbf{H}^{\rm H} \mbf{v} \right\}
- (\mbf{v}^{t-1})^{\rm H} \mbf{H} \mbf{H}^{\rm H} \mbf{v}^{t-1}, \label{Talor}
\end{align}
which gives an approximation of $f(\mbf{v})$ and the equality holds at point $\mbf{v} = \mbf{v}^{t-1}$.
By maximizing this approximation subject to the constraint \eqref{P1-vn}, the closed-form solution at the $t$-th iteration is readily given by
\begin{align}
\mbf{v}^{t} =  e^{\jmath \, \mathrm{arg} (\mbf{H} \mbf{H}^{\rm H} \mbf{v}^{t-1})}. \label{v_update}
\end{align}

Based on the above discussions, we provide the implementation details
of the HPB-SPP algorithm in Algorithm \ref{alg:SCA}.

\section{Numerical Results}
In this section, we use simulation results to evaluate the proposed HPB design.
Due to practical restrictions such as topography, RISs may not be always deployed at the BS-side or the user-side as in \cite{You2020Deployment} to alleviate the product-distance path-loss effect to the best extent.  
In this letter, we consider the worst case that the distances from the BS to the $n$-th RIS and from the $n$-th RIS to the user, denoted by $d_{1,n}$ and $d_{2,n}$ respectively, are equally set to $50$ m.
The BS is equipped with $M=8$ antennas, emitting signals with carrier wavelength $\lambda = 0.1$ m.
Without loss of generality, $D_n$ and $K_n$ are assumed to be the same and denoted by $P$.
The AoAs/AoDs are Laplacian distributed with an angular spread $\sigma_{\rm AS} = 10^{\circ}$.
The complex path gain $\alpha_{n,d}$ of the BS-RIS channel is circularly symmetric complex Gaussian (CSCG) distributed, i.e., $\alpha_{n,d} \sim \mathcal{CN}(0,{\sigma}_{n,d}^{2})$, where ${\sigma}_{n,d}^{2}$s are randomly generated form an exponential distribution and normalized such that $\sum_{d=1}^{D_n} {\sigma}_{n,d}^{2} = 1$.
The complex path gain $\beta_{n,k}$ of the RIS-user channel is similarly modeled and here we omit the details.
Additionally, the path-loss coefficient of the cascaded channel is given by \cite[Propersition 1]{WankaiTang2020TWC_PathLoss}
\begin{align}
PL_n = G_{BS} G_{RIS,n} G_{User} \frac{\Delta^2 L^4 \lambda^4}{64 \pi^3 d_{1,n}^2 d_{2,n}^2 }, \label{WankaiTang-PathLoss}
\end{align}
where $G_{BS} = 5$ dBi, $G_{RIS,n} = 5$ dBi, and $G_{User} = 0$ dBi are the antenna gains at the BS, the $n$-th RIS, and the user, respectively.
The remaining parameters are set as follows: $ \Delta = \frac{1}{2}$, $ p = 0.01$ W, $\sigma^2 = -100$ dBm, $I_{\rm sca} = 10^3$, $\epsilon_{\rm sca} = 10^{-6}$.
The experiments are carried out on a Windows x64 machine with $3.4$ GHz CPU and $256$ GB RAM.
All the simulation curves are obtained by averaging $1000$ independent channel realizations.
The following PB algorithm and other comparison schemes are considered.
\begin{itemize}
	\item \textbf{PB-SCA:} the SCA method is adopted to solve the traditional PB problem, where the closed-form expressions can be derived in each iteration.
	\item \textbf{HPB-AO:} the simulated annealing (SA) algorithm and the SCA method is adopted to alternatingly optimize $\mathbf{Q}$ and $\mathbf{v}$, respectively.
	\item \textbf{HPB-ES:} for the single-RIS setup, i.e., $N=1$, a $400 \times 400$ gird is established to exhaustively search (ES) the best solutions of $q_1^x$ and $q_1^y$.
	\item \textbf{Random Phase Shifts:} for each channel realization, the result is averaged with $1000$ randomly generated reflection coefficients.
\end{itemize}

\begin{figure}
	\centering
	\includegraphics[width=1\columnwidth]{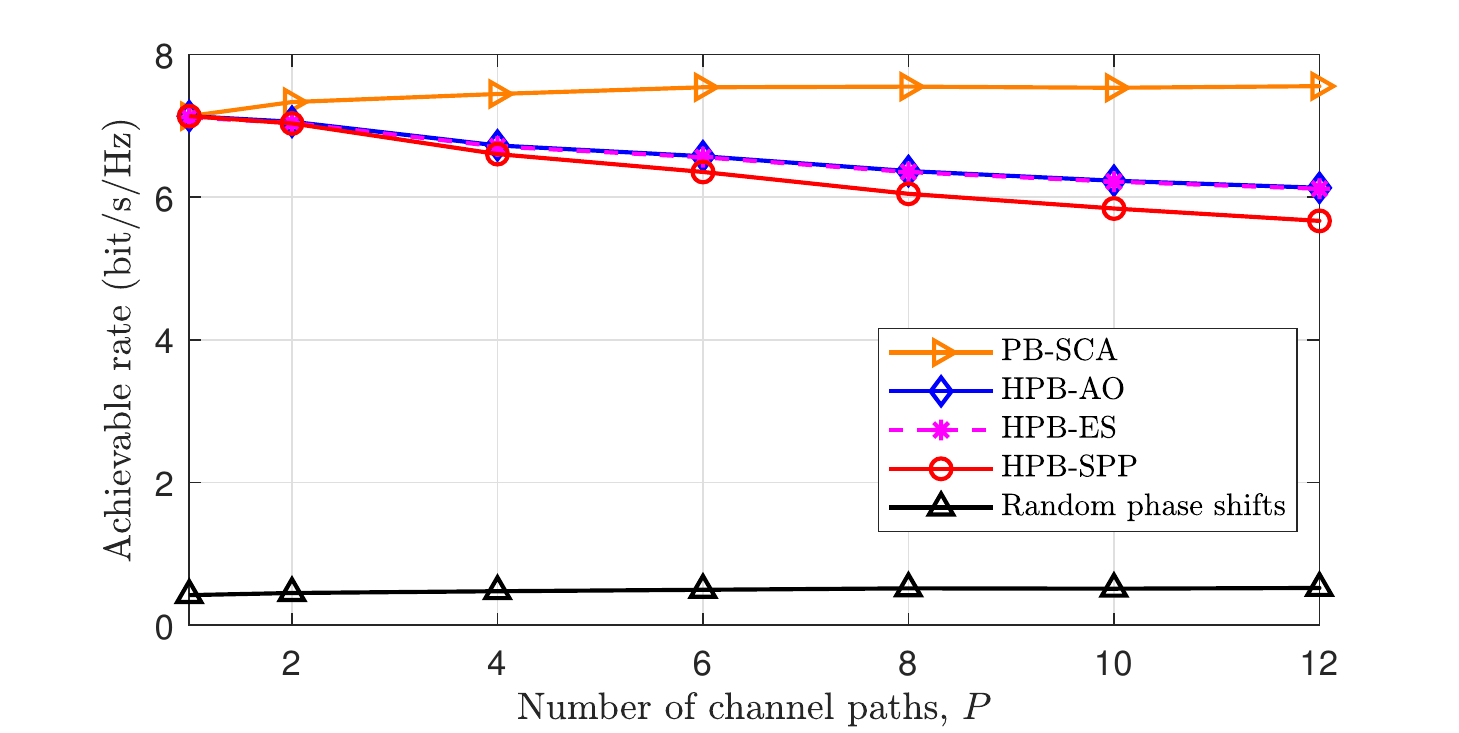}
	\vspace{-1.4em}
	\caption{Achievable rate versus $P$, with $N=1$ and $L^2 = 900$.}
	\label{Fig_Path}
	\vspace{-.6em}
\end{figure}

Fig. \ref{Fig_Path} considers the single-RIS setup and plots the achievable rate of the user w.r.t. the number of channel paths $P$, with $L^2 = 900$.
It is observed that the performance of the three HPB algorithms, namely HPB-AO, HPB-ES, and HPB-SPP, are far beyond the random phase shifts settings. 
Also, the performance gap between the three HPB algorithms and PB-SCA is acceptable small.
For $P=1$, the solutions of the HPB algorithms overlap with that obtained by PB-SCA.
Moreover, 
compared to HPB-ES and HPB-AO,
the performance loss of HPB-SPP is marginal, while the former two algorithms incur unaffordable complexity especially when $N$ is large.

\begin{figure}
	\centering
	\includegraphics[width=1\columnwidth]{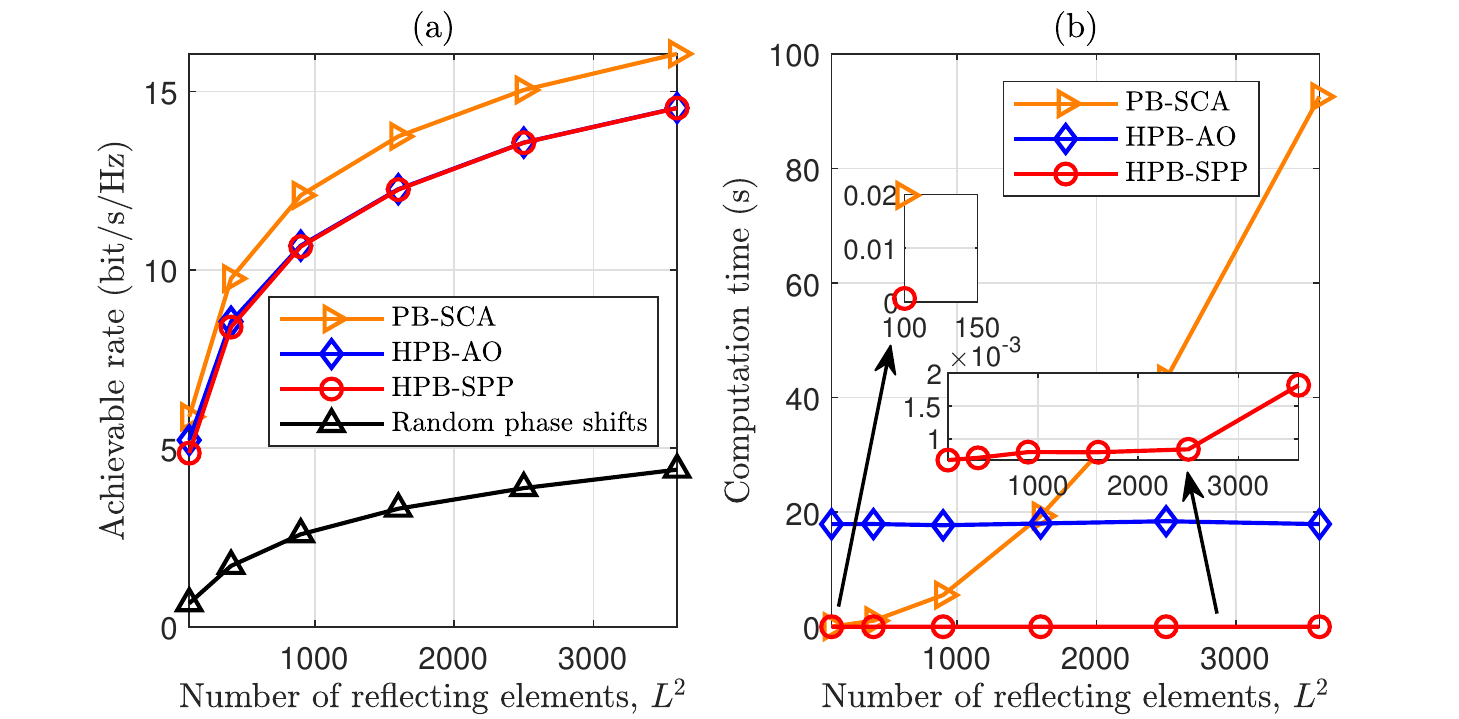}
	\vspace{-1.4em}
	\caption{(a) Achievable rate and (b) computation time versus $L^2$, with $N=3$ and $P=8$.}
	\label{Fig_Element}
	\vspace{-.6em}
\end{figure}

In Fig. \ref{Fig_Element}, the achievable rate and computation time are respectively evaluated against the number of reflecting elements $L^2$ when $N=3$ and $P=8$.
It is seen that the achievable rate increases as $L^2$ becomes large for all the four curves.
Compared with HPB-SPP, as $L^2$ increases, the achievable rate improvement of HPB-AO decreases since the performance of the SA procedure in HPB-AO is sensitive to the dimension of variable space.
We remark that although the PB design performs slightly better than the HPB designs, their computation complexities are quite different.
It is seen from Fig. \ref{Fig_Element}(b) that the computation time of PB-SCA rises very fast as $N$ increases, and requires about $100$ seconds to obtain a PB solution when $L^2=3600$.
However, the computation time of HPB-SPP is extremely low, and it is easy to meet the real-time signal processing requirement (less than $0.002$ second even when $L^2=3600$).
Moreover, increasing the number of reflecting elements does not significantly influence the computation time of the HPB algorithms.
The reason is that in HPB design, the number of variables related to RIS configuration remains constant when varying $L$.

\begin{figure}[t]
	\centering
	\includegraphics[width=1\columnwidth]{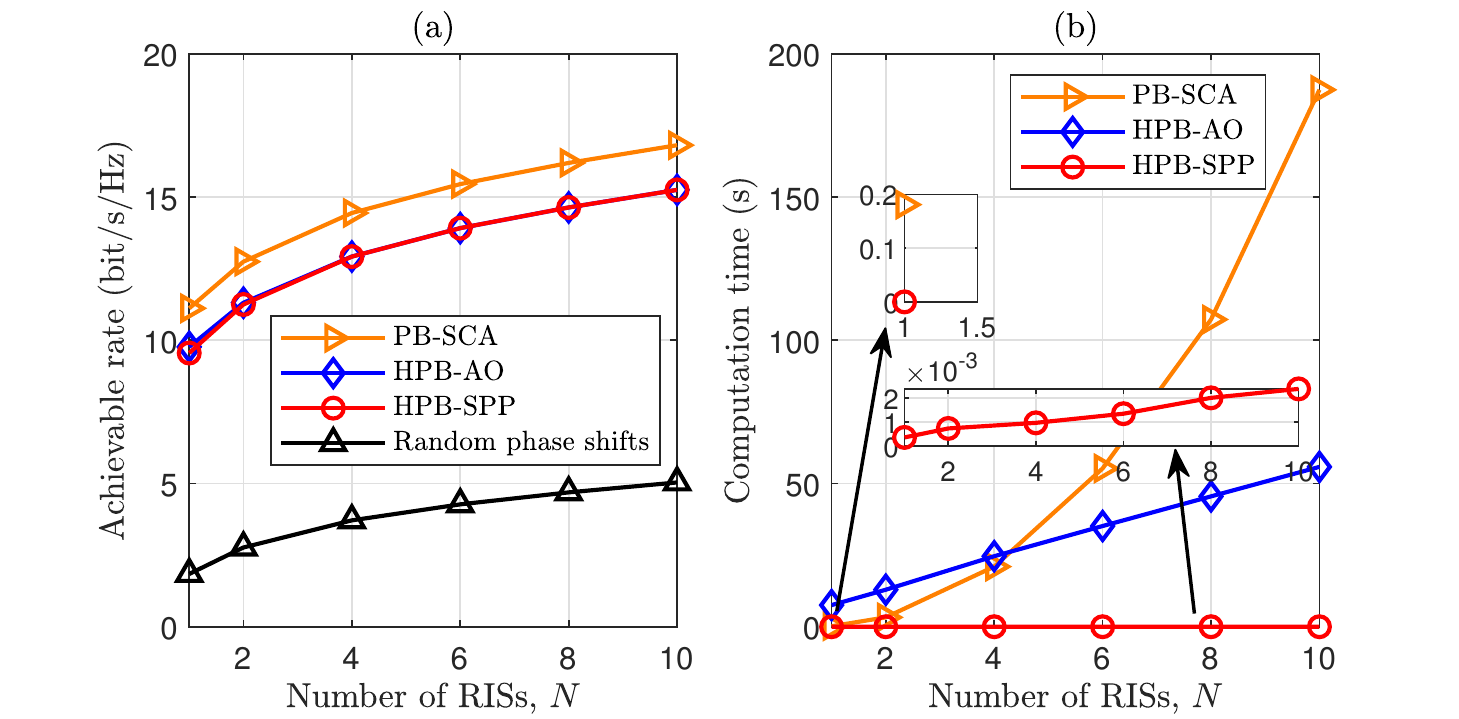}
	\vspace{-1.4em}
	\caption{(a) Achievable rate and (b) computation time versus $N$, with $L^2=1600$ and $P=8$.}
	\label{Fig_RIS}
	\vspace{-.6em}
\end{figure}

The performance and complexity comparisons are also made in Fig. \ref{Fig_RIS} by varying the number of RISs. 
The major difference is that the computation time of the HPB algorithm increases as $N$ increases,
since the dimension of the variable space for HPB design scales with $N$, i.e., $3N$ for URA-shaped RISs.

\section{Conclusion}
This letter proposed a new concept, i.e., HPB, for RIS reflecting coefficients design by assuming arithmetic-sequence-structured phase shifts.
Experimental results evidence that the proposed solution achieves a close-to-ideal performance with significantly reduced computational complexity in PB design.

\bibliographystyle{IEEEtran}
\bibliography{IEEEabrv,mybib}

\begin{thebibliography}{10}
\providecommand{\url}[1]{#1}
\csname url@samestyle\endcsname
\providecommand{\newblock}{\relax}
\providecommand{\bibinfo}[2]{#2}
\providecommand{\BIBentrySTDinterwordspacing}{\spaceskip=0pt\relax}
\providecommand{\BIBentryALTinterwordstretchfactor}{4}
\providecommand{\BIBentryALTinterwordspacing}{\spaceskip=\fontdimen2\font plus
\BIBentryALTinterwordstretchfactor\fontdimen3\font minus
  \fontdimen4\font\relax}
\providecommand{\BIBforeignlanguage}[2]{{%
\expandafter\ifx\csname l@#1\endcsname\relax
\typeout{** WARNING: IEEEtran.bst: No hyphenation pattern has been}%
\typeout{** loaded for the language `#1'. Using the pattern for}%
\typeout{** the default language instead.}%
\else
\language=\csname l@#1\endcsname
\fi
#2}}
\providecommand{\BIBdecl}{\relax}
\BIBdecl

\bibitem{ChongwenHuang2019TWC}
C.~{Huang} \emph{et~al.}, ``Reconfigurable intelligent surfaces for energy
  efficiency in wireless communication,'' \emph{IEEE Trans. Wireless Commun.},
  vol.~18, no.~8, pp. 4157--4170, 2019.

\bibitem{QingqingWu2019TWC}
Q.~{Wu} and R.~{Zhang}, ``Intelligent reflecting surface enhanced wireless
  network via joint active and passive beamforming,'' \emph{IEEE Trans.
  Wireless Commun.}, vol.~18, no.~11, pp. 5394--5409, 2019.

\bibitem{He2019Cascaded}
Z.~{He} and X.~{Yuan}, ``Cascaded channel estimation for large intelligent
  metasurface assisted massive {MIMO},'' \emph{IEEE Wireless Commun. Lett.},
  vol.~9, no.~2, pp. 210--214, 2020.

\bibitem{CunhuaPan2019TWC}
C.~{Pan} \emph{et~al.}, ``Multicell {MIMO} communications relying on
  intelligent reflecting surfaces,'' \emph{IEEE Trans. Wireless Commun.},
  vol.~19, no.~8, pp. 5218--5233, 2020.

\bibitem{HuayanGuo2020WSR}
H.~{Guo} \emph{et~al.}, ``Weighted sum-rate maximization for reconfigurable
  intelligent surface aided wireless networks,'' \emph{IEEE Trans. Wireless
  Commun.}, vol.~19, no.~5, pp. 3064--3076, 2020.

\bibitem{Ning2020SPGM}
B.~{Ning} \emph{et~al.}, ``Beamforming optimization for intelligent reflecting
  surface assisted {MIMO}: A sum-path-gain maximization approach,'' \emph{IEEE
  Wireless Commun. Lett.}, vol.~9, no.~7, pp. 1105--1109, 2020.

\bibitem{Nanfang2011Snell}
N.~Yu \emph{et~al.}, ``Light propagation with phase discontinuities:
  generalized laws of reflection and refraction,'' \emph{Science}, vol. 334,
  no. 6054, pp. 333--337, 2011.

\bibitem{Larsson2020Physics_WCL}
{\"O}.~{\"O}zdogan, E.~Bj{\"o}rnson, and E.~G. Larsson, ``Intelligent
  reflecting surfaces: Physics, propagation, and pathloss modeling,''
  \emph{IEEE Wireless Commun. Lett.}, vol.~9, no.~5, pp. 581--585, 2019.

\bibitem{3GPP}
3GPP, ``Study on channel model for frequencies from 0.5 to 100 {GH}z (3{GPP}
  {TR} 38.901 version 16.1.0 release 16),''
  \emph{https://www.3gpp.org/ftp/Specs/archive/38\_series/38.901/38901-g10.zip},
  2019.

\bibitem{JiguangHe2020ANM_Hybrid}
R.~Schroeder, J.~He, and M.~Juntti, ``Passive {RIS} vs. hybrid {RIS}: A
  comparative study on channel estimation,'' \emph{arXiv:2010.06981}, 2020.

\bibitem{You2021DoubleIRS}
C.~{You}, B.~{Zheng}, and R.~{Zhang}, ``Wireless communication via double
  {IRS}: Channel estimation and passive beamforming designs,'' \emph{IEEE
  Wireless Commun. Lett.}, vol.~10, no.~2, pp. 431--435, 2021.

\bibitem{You2020Deployment}
C.~You, B.~Zheng, and R.~Zhang, ``How to deploy intelligent reflecting surfaces
  in wireless network: {BS}-side, user-side, or both sides?'' \emph{arXiv
  preprint arXiv:2012.03403}, 2020.

\bibitem{WankaiTang2020TWC_PathLoss}
W.~{Tang} \emph{et~al.}, ``Wireless communications with reconfigurable
  intelligent surface: Path loss modeling and experimental measurement,''
  \emph{IEEE Trans. Wireless Commun.}, pp. 1--1, 2020.

\end{thebibliography}

\end{document}